\documentclass[prb,showpacs,showkeys,preprintnumbers,amsmath,amssymb]{revtex4}
\usepackage{graphicx}
\usepackage{dcolumn}
\usepackage{bm}

\begin{document}

\title{Implementation Schemes for the Factorized Quantum Lattice-Gas Algorithm
for the One Dimensional Diffusion Equation using
Persistent-Current Qubits}

\author{David M. Berns}
 \email{dmb@mit.edu}
 \affiliation{Department of Physics,\\
Massachusetts Institute of Technology,\\ Cambridge, Massachusetts
02139}
\author{T. P. Orlando}
 \affiliation{Department of Electrical Engineering and Computer
 Science,\\
Massachusetts Institute of Technology,\\ Cambridge, Massachusetts
02139}

\date{\today}

\begin{abstract}
We present two experimental schemes that can be used to implement
the Factorized Quantum Lattice-Gas Algorithm for the 1D Diffusion
Equation with Persistent-Current Qubits. One scheme involves
biasing the PC Qubit at multiple flux bias points throughout the
course of the algorithm. An implementation analogous to that done
in Nuclear Magnetic Resonance Quantum Computing is also developed.
Errors due to a few key approximations utilized are discussed and
differences between the PC Qubit and NMR systems are highlighted.
\end{abstract}


\maketitle

\section{Introduction}

Most algorithms designed for quantum computers will not best their
classical counterparts until they are implemented with thousands
of qubits.  For example, the factoring of binary numbers with a
quantum computer is estimated to be faster than a classical
computer only when the length of the number is greater than about
500 digits\cite{Karl}. Accounting for error correction
circuitry\cite{Chuang} would bring the size of the needed quantum
computer to be in the thousands of qubits. In contrast, the
Factorized Quantum Lattice-Gas Algorithm (FQLGA)\cite{Yepez} for
fluid dynamics simulation, even when run on a quantum computer
significantly smaller than the one just discussed, has significant
advantages over its classical counterparts.

The FQLGA is the quantum version of classical lattice-gases
(CLG)\cite{Dieter}. CLG are an extension of classical cellular
automata with the goal of simulating fluid dynamics without
reference to specific microscopic interactions. The binary nature
of the CLG lattice variables is replaced for the FQLGA by the
Hilbert space of a two-level quantum system. The results of this
replacement are similar to that of the lattice-Boltzmann model,
but with a few significant differences\cite{Eff}. The first is the
exponential decrease in required memory.  The second is the
ability to simulate arbitrarily small viscosities.


As of today there is a plethora of qubits to choose from when
designing a quantum computer, and a promising class is
superconducting qubits based on Josepshon junction
circuits\cite{PC, RF, Phase, Charge, Hybrid}. One major advantage
of any of these superconducting systems is the ability to
precisely engineer the quantum Hamiltonian, which extends from
single qubit design to multi-qubit coupling arrangements to
measurement engineering. The quantum computer considered here will
be built using the Persistent-Current Qubit (PC Qubit)\cite{PC}.

The goal of this paper is to show how one can implement a one
dimensional version of the FQLGA with the PC Qubit.  To this end
we will begin by reviewing the algorithm, specifically the one
that simulates the diffusion equation, without a loss of
generality in understanding the essence of the algorithm or its
general requirements. We will then review the PC qubit and show
explicitly how to implement the algorithm with this system. Some
important differences between the PC qubit and the two-state
system studied in Nuclear Magnetic Resonance Quantum Computation
(NMRQC)\cite{NMR} will be shown to allow for some interesting new
techniques in implementing quantum logic. We will also show how to
implement the algorithm with the PC qubit in a very analogous way
to NMRQC schemes\cite{NMRIMP}, with a few significant differences.

\section{FQLGA for the 1D Diffusion Equation}
\label{algorithm}

The first thing one must do in the FQLGA is to define a lattice.
Each lattice point $\vec{n}$ will represent a unique position in
the simulated fluid. The simulation will contain a finite number
of lattice points, hence space is discretized in the simulation.

Next one must encode the mass density $\rho$ of the fluid at each
lattice site. In the FQLGA this is done by building at each
lattice site a set $\{i\}$ of coupled qubits. Each qubit
represents the motion of particles on the microscopic level in one
of a finite set of directions. For the diffusion equation in one
dimension, at any point in your fluid, there are only two possible
directions for each particle to be moving, to the left and to the
right. Hence, only two qubits are needed to specify the mass
density ${\rho}^n$ at each lattice site. This intuitive reasoning
does not extrapolate to higher dimensional simulations because
even in two dimensions there would be an infinite number of
directions particles could travel in. In higher dimensions one
must adhere to much more mathematical conditions to decide on the
small set of directions one must include for a faithful
simulation\cite{Dieter}. The probability P of a particle to be
participating in the motion assigned to each qubit will be encoded
in the probability amplitude of the qubit being in its excited
state $|1\rangle$. The state of a qubit is thus set to
     \begin{equation}
    {|\Psi_i}^n\rangle=\sqrt{1-{P_i}^n}|0\rangle+\sqrt{{P_i}^n}|1\rangle
    \end{equation}
where $i$ is the qubit index, $n$ is the lattice site index, and
$|0\rangle$ is the ground state of the qubit. For the one
dimensional problem considered here, $i=\{1,2\}$ and $n=\{1,N\}$
where N is the number of lattice sites used in the simulation. One
can easily conceive of fluids of multiple phases with multiple
types of interactions even in one dimension, in which the size of
$\{i\}$ would be much larger, but this will not be considered
here. The mass density $\rho$ is then calculated by summing the
occupation probabilities for all qubits at a node.  At time t=0 in
a 1D simulation the occupation probabilities ${P_1}^n$ and
${P_2}^n$ are set to ${\rho}^n/2$, which is the condition for
local equilibrium in the fluid\cite{Diff}.

Now that the fluid is initialized, one must account for the
interaction of particles in the fluid. These collisions are
encoded by the application of a unitary transformation to the
coupled systems at each lattice site.  For the 1D diffusion
equation this unitary transformation is
 \begin{equation}
\sqrt{swap}=\frac{1}{2}
\begin{pmatrix}
  2 & 0 & 0 & 0 \\
  0 & 1+i & 1-i & 0 \\
  0 & 1-i & 1+i & 0 \\
  0 & 0 & 0 & 2 \\
\end{pmatrix}.
 \end{equation}

The basis for computation is the set of four product states:
$|0\rangle$$|0\rangle$, representing no particles at the site,
$|0\rangle$$|1\rangle$, representing the existence of only a
particle moving to the right at the site, $|1\rangle$$|0\rangle$,
representing the existence of only a particle moving to the left
at the site, and $|1\rangle$$|1\rangle$, representing particles
moving in both directions at the site. To conserve particle number
there can be coupling only between the middle two states. The
identity transformation on the first and last states corresponds
to no collisions and a perfectly elastic collision respectively.
Transformation of the middle two states was something that never
existed in the classical algorithm because there was no
superposition of these two states.

After collision the states of the qubits at each lattice site are
in general entangled, and we denote that state as
$|$\textbf{${\Upsilon}^n$}$\rangle$. The state of each qubit is
then measured, denoted by \textbf{$|{\chi_i}^n\rangle$}, and the
process described thus far is repeated many times to achieve an
ensemble average.
Upon completion of these measurements one will have found the
post-collision outgoing occupation probabilities, denoted by
${P_i}^n$ once again. Note that the occupation probabilities now
represent something very different than before the collision. The
particles have now interacted and are ready to move to the next
lattice site.

One must now ``stream" the occupation probabilities to their new
lattice sites.  This is done in a classical computer by storing
the occupation probabilities at each lattice site that are coming
from adjacent lattice sites due to collisions. More precisely,
${{P_1}^n}$ becomes ${{P_1}^{n+1}}$ and ${{P_2}^n}$ becomes
${{P_2}^{n-1}}$. Periodic boundary conditions are assumed when
streaming at the edges of the fluid.

To find the mass density ${\rho}^n$ at t=1 one simply adds the
occupation probability for both qubits at site $n$ once streaming
has been done.  One time step of the algorithm has now been
completed. To simulate the next time step simply start the above
procedure all over again except now setting the initial states
with the new occupation probabilities just found.

\begin{figure}
\includegraphics[height=3.0in]{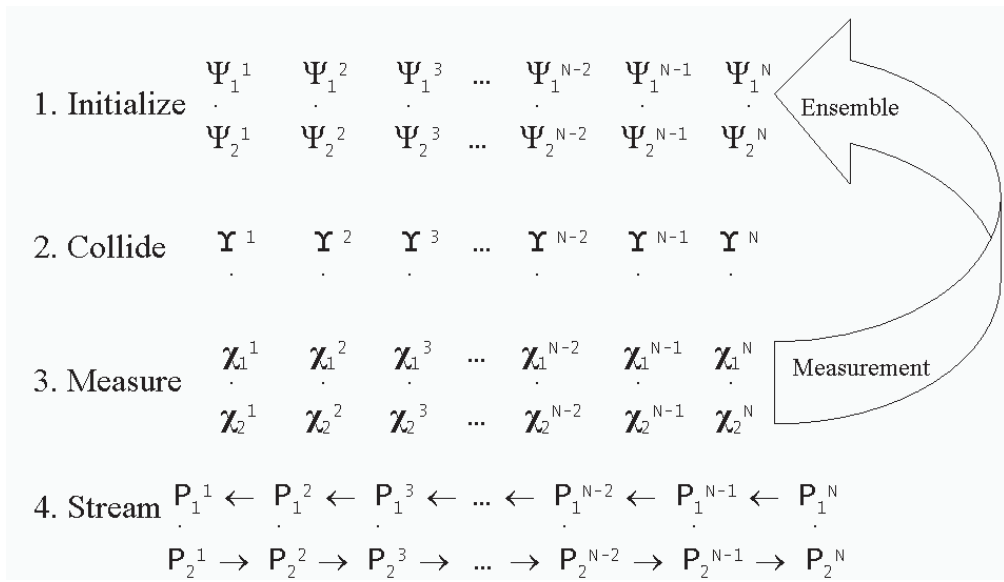}
\caption{General summary of the four major steps that comprise one
time step of the 1D FQLGA fluid dynamics simulation. The sequence
of initialization of mass density, collision of particles, and
measurement of post-collision states is repeated many times to
make an ensemble measurement. Propagation between collisions is
accomplished by storing the adjacent occupation probabilities for
a given site in a classical computer, where the mass density is
then calculated for this time step. Subsequent time steps utilize
these ``streamed'' occupations when initializing again for the
next set of collisions and ``streaming''.} \label{FQLGA}
\end{figure}

The algorithm can be summarized by four major steps, which are
illustrated in figure \ref{FQLGA}.  The first step encodes the
initial state of the fluid by quantum mechanically setting the
state $|{\Psi_i}^n\rangle$ of each qubit at each lattice site. The
second step transforms the two-qubit product state at each lattice
site to in general an entangled state, whose state is denoted by
\textbf{$|\Upsilon^n\rangle$}. Third one makes a projective
measurement of the post-collision states
\textbf{$|{\chi_i}^n\rangle$}, and one must repeat the first three
steps to find the outgoing occupation probabilities ${P_i}^n$. In
the fourth and final step one streams the mass density with the
appropriate post-collision occupation probabilities, from the left
with particles representing positive momentum, and from the right
with particles representing negative momentum, and the mass
density is calculated. Subsequent time steps are identical except
for a change in the initial mass density profile, i.e., initial
qubit states in the first step.

\section{Persistent-Current Qubit}\label{qubit}

\begin{figure}
\includegraphics[height=2.5in]{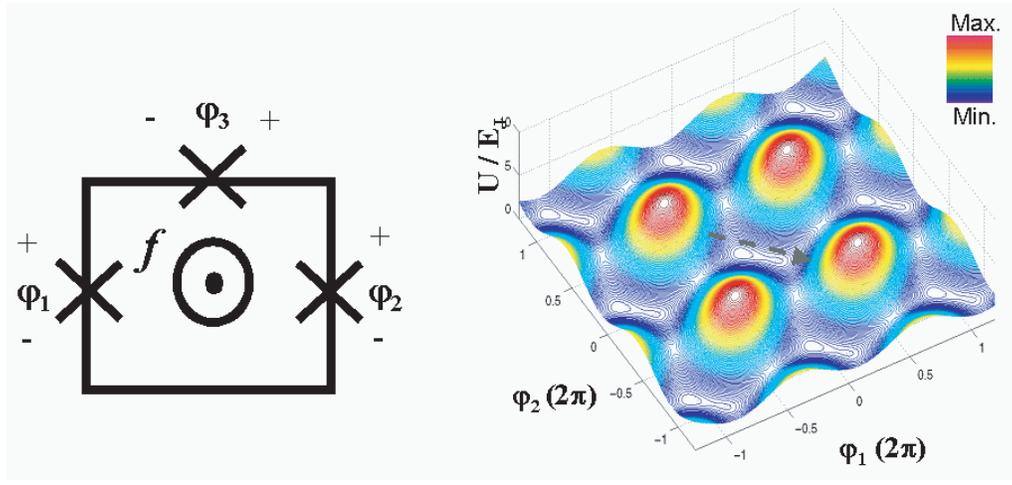}
\caption{(a) Schematic drawing of the PC Qubit.  The x's represent
Josephson junctions, with all connecting leads made of the same
superconductor that is part of the junctions. The sign conventions
chosen when summing phases are shown, and the magnetic flux
penetrating the loop (in units of $\Phi_o$) is labeled by $f$. (b)
The potential energy of the full Hamiltonian for the PC Qubit is
plotted when the system is biased at $f=0.495\Phi_o$. The phase
particle sees an infinite 2D lattice with unit cells resembling a
double well potential.} \label{DiagANDPot}
\end{figure}

The fundamental unit of quantum logic we will use to implement the
algorithm is the PC Qubit\cite{PRB}.  It consists of a
superconducting loop that is interrupted by three Josephson
junctions, pictured as x's in figure \ref{DiagANDPot}(a).  The
magnetic flux $\Phi$ is the only control field for our qubit, and
as shown in the figure, is usually denoted by $f=\Phi/\Phi_o$
where $\Phi_o = h/2e$ is a single flux quantum, $h$ is Planck's
constant, and $e$ is the magnitude of the charge of an electron.
Physically, a Josephson junction is a small layer of insulator
sandwiched between superconductors, so our system is a
superconducting loop interrupted by three layers of insulator
about 1nm thick. For single qubit manipulation the magnetic flux
through the loop will be modified. The flux seen by a DC SQUID
magnetometer, a combination of applied flux and qubit-induced
flux, will serve as our measurement variable.

The Hamiltonian of the qubit is derived by considering a circuit
element model of our system, which consists of three Josepshon
junctions, where two junctions have the same cross-sectional area,
and the third is smaller by a factor of $\alpha$.  The constituent
relations for an ideal Josephson junction are
     \begin{subequations}
     \begin{align}
      I&=I_c\sin(\varphi)\label{Ijj} \\
      V&=\frac{\Phi_o}{2\pi}\frac{d\varphi
}{dt}\label{Vjj}
      \end{align}
    \end{subequations}
where $I$ is the current through the junction, $V$ is the voltage
across the junction, $I_c$ is the maximum current the junction can
hold without a voltage appearing across it,
$\varphi$=$\theta_1$-$\theta_2$, and $\theta_{1,2}$ is the phase
of the plane wave macroscopic wavefunction that characterizes the
superconductor condensate on the +,- side of the junction
respectively. Note that $I_c$ is a function linear in the
cross-sectional area of the junction, and hence the third junction
has a lower $I_c$ by a factor of $\alpha$.

The energy associated with an ideal Josephson junction is found by
integrating the power from time $t=0$ to some final time $t_o$,
which is equivalent to an integral from zero phase to some phase
$\varphi$. The energy it takes to set the phase of a Josephson
junction to $\varphi$ is
\begin{equation}
E=\int^{t_o}_o(I_c\sin\varphi ')(\frac{\Phi_o}{2\pi}\frac{d\varphi
'}{dt})dt=\frac{\Phi_o I_c}{2\pi} \int^{\varphi}_o \sin\varphi'
d\phi'=E_j(1-\cos\varphi)
\end{equation}
where $E_j=\Phi_o I_c/2\pi$.

By including the charging energies due to the capacitance of the
junctions, the Hamiltonian of our circuit is\cite{PRB}
     \begin{equation}
      H=\frac{P_p^2}{2M_p}+\frac{P_m^2}{2M_m}+E_j[2+\alpha-2\cos(\varphi_p)\cos(\varphi_m)-\alpha\cos(2 \pi f
      +2\varphi_m)]\label{H}
    \end{equation}
    where $\varphi_p = \varphi_1+\varphi_2$ , $\varphi_m =
    \varphi_1-\varphi_2$, $P_p=M_p d\varphi_p/dt$, $P_m=M_m d\varphi_m/dt$,
    $M_p=(\Phi_o/2\pi)^2 2C$, and $M_m=(\Phi_o/2\pi)^2 2C(1+2\alpha)$.
The number of degrees of freedom in the problem was reduced by the
fluxoid quantization condition\cite{Terry}
\begin{equation}
\varphi_1+\varphi_3-\varphi_2 = 2\pi n + \frac{2\pi\Phi}{\Phi_o}
\end{equation}
which forces the sum of the gauge invariant phases to be
proportional to the amount of flux quanta modulo an integer
multiple of $2\pi$.

We have chosen to associate the capacitive energy, the first two
terms in (\ref{H}), with the kinetic energy, and the ideal
Josephson energy, the last four terms in (\ref{H}), with the
potential energy. The potential energy is that of an infinite
lattice of double wells, as seen in figure \ref{DiagANDPot}(b).
The arrow in the plot shows the direction one would take to
traverse from one side of a double well to another. The barrier
between the left and right sides of a single double well is much
lower than any barrier between different double wells.

Though quantum mechanics plays a foremost role in deriving the
constitutive relations for the superconducting circuit elements,
the Hamiltonian for the circuit itself so far has been classical.
The quantum version of the circuit can be understood by imagining
a phase ``particle" in the potential shown in figure
\ref{DiagANDPot}(b). The behavior of this ``particle" is analogous
to a particle with an anisotropic mass moving in a 2D periodic
potential, and so there exist energy bands in a $\vec{k}$-space,
which is here related to the charge stored capacitively by the
Josephson junctions. By properly choosing $E_j/E_c$, where
$E_c=e^2/2C$, one can remove any $\vec{k}$ (and hence charge)
dependence in the energy of the system, and hence can reduce the
problem to that of an effective double well. What we have done is
choose parameters such that tunneling between adjacent double
wells can be neglected relative to the tunneling within a double
well in the tight-binding solution, making the solution
effectively that of a single double well.

\begin{figure}
\includegraphics[height=3.0in]{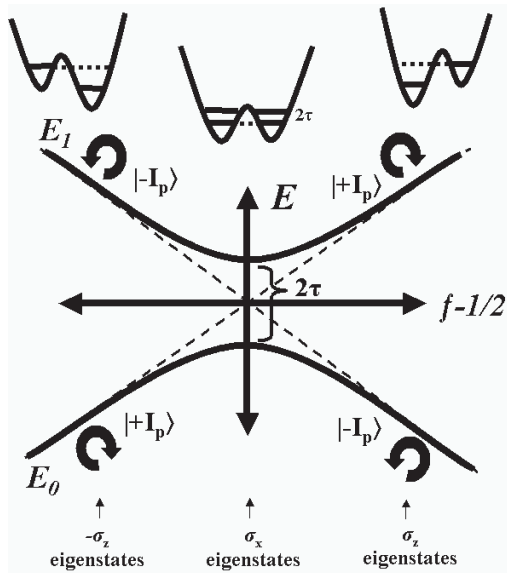}
\caption{The energy levels of the PC Qubit are shown as a function
of $f$.  The eigenstates of the system change with $f$ and are
labeled on the plot.  The change in the potential of the phase
particle is also depicted at the top of the plot.  The energy
difference between the states at $f=1/2$ is seen to be twice the
intra-well tunnelling.} \label{TwoLevel}
\end{figure}

By considering only the lowest two levels of the double well, the
equivalent Hamiltonian is
     \begin{equation}
      \hat{H}=\Phi_oI_p(\emph{f}-\frac{1}{2})\hat{\sigma}_z-\tau\hat{\sigma}_x
    \end{equation}
where $\pm$$I_p$ are the eigenvalues of circulating current for
the two $\hat{\sigma}_z$ eigenstates and $\tau$ is the tunneling
element from one side of the double well to the other. The
energies of the two eigenstates along with a sketch of the double
well as a function of applied flux are shown in figure
\ref{TwoLevel}. One significant difference between this qubit and
the one used in NMRQC is the presence of the $\hat{\sigma}_x$
term. The implication of such a term is that the energy of the
eigenstates as well as the eigenstates themselves change as the
bias field is modified.  In figure \ref{TwoLevel} we see that at
the classical degeneracy point $f=1/2$ the qubit's eigenstates are
$\hat{\sigma}_x$ eigenstates, while far from $f=1/2$, but still
far from $f=1$, the eigenstates are those of $\hat{\sigma}_z$. The
same thing happens for $f<1/2$, but now the eigenstates have
switched energies, i.e., the ground state here is the first
excited state of $\hat{\sigma}_z$ and vice versa.

The PC Qubit has some advantages over other superconducting
qubits.  Charge fluctuations, a consequence of trapped substrate
charge, are deemed inconsequential through the choice of
parameters used when designing the PC Qubit circuit. Also, flux
noise has been reduced in this system over other flux qubits since
this system has a smaller loop.

A typical conceptual misconception can be addressed at this point.
The two different states used in computation are not related to
single Cooper pair behavior.  Rather, they are macroscopically
distinct states described by the circulating current due to
millions of Cooper pairs, characterized by different average
induced fluxes when in a magnetic field.

As seen in section \ref{algorithm}, the qubits will need to be
coupled.  For the PC Qubit, just as microwaves can only be coupled
in through $\hat{\sigma}_z$, coupling between qubits can only be
of the form $\hat{\sigma}_z \hat{\sigma}_z$. Other coupling terms
can be introduced by design, even for these planar
devices\cite{PRB}.

\section{Implementation with the PC Qubit}

We now show how one can use the PC Qubit to simulate the 1D
diffusion equation.  In section \ref{imp1} we elaborate on a
scheme based upon changing the flux bias points of the qubits
during the algorithm, which will lead to a very general
initialization scheme, but a less general collision. In section
\ref{imp2} we discuss a more general collision, analogous to that
done in NMRQC, and how to initialize the qubits before this
general collision.

\subsection{The Multiple Bias Point Implementation}\label{imp1}

The first of the four steps of the algorithm is initializing each
qubit at each node.  As discussed in section \ref{algorithm}, each
qubit must be initialized into a state of real and positive phase
in its own Hilbert space. This set of states consists of all those
lying on the real phase geodesic between the ground and first
excited states on the Bloch sphere. The ground state of the PC
Qubit as a function of applied flux coincidentally also occupies
exactly this geodesic on the Bloch sphere, as discussed in section
\ref{qubit}. Initialization can thus be accomplished while staying
in the ground state by adiabatically changing the applied magnetic
flux, as depicted in figure \ref{TwoLevel}.

The flux used to set the state of one qubit will be affected by
the state of the other qubit and its bias current.  This permanent
inductive coupling can be accounted for by slightly adjusting the
applied flux to compensate for the flux introduced by the other
qubit and its bias line. All errors due to approximations made
when initializing by rotating dynamically on the Bloch sphere,
including errors due to decoupling, have been
avoided\cite{NMRIMP}. We emphasize that the initialization portion
of the algorithm is identical for any simulation, whether it be
for a different equation, a multi-phase simulation, or in a
different number of dimensions.

The second step of the algorithm is the collision.  Here we study
a very specific unitary transformation, the $\sqrt{swap}$
described in section \ref{algorithm}.  This matrix simply
``half-way" swaps the middle two (first and second excited)
computational states of the coupled system. In NMRQC, the coupled
eigenstates are exactly those computational states, but there are
no direct matrix elements connecting these states\cite{Berman}.
When the PC Qubits are coupled, the first and second excited
states of the four-level system, denoted as $|1\rangle$ and
$|2\rangle$ respectively, are in general not the same as the
computational basis states the $\sqrt{swap}$ intends to affect.
However, the dc bias fields of each qubit can be tuned to make
these two sets of eigenstates coincide. Once this is done, one can
then implement the $\sqrt{swap}$ by simply oscillating the
magnetic field bias at the frequency corresponding to the energy
difference between the middle two eigenstates.  This is just a
Rabi oscillation between the middle two eigenstates, and since one
wants to only ``half-way" swap the states, the radiation should
only be left on for a quarter of a Rabi period.

\begin{figure}
\includegraphics[height=3.0in]{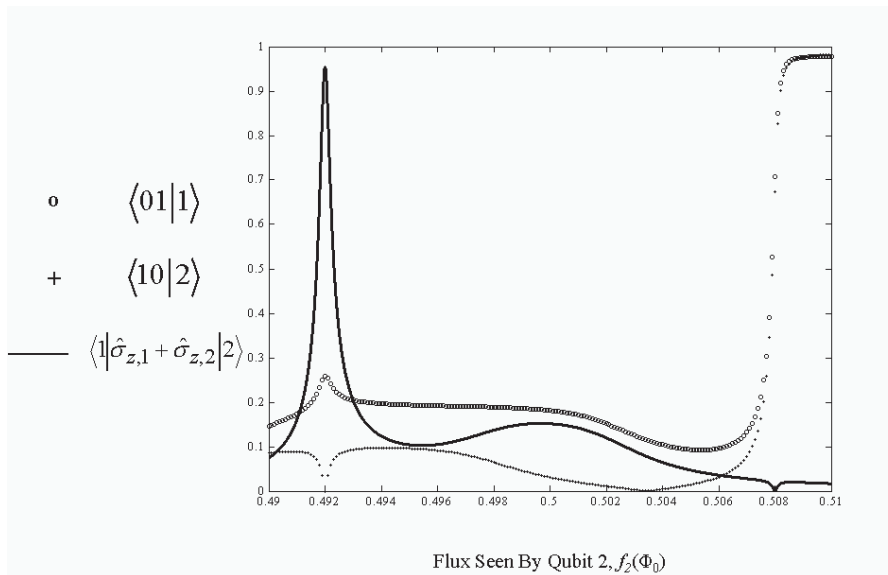}
\caption{The overlap between the first (second) excited state
$|1\rangle$ ($|2\rangle$) of the PC Qubit coupled system and the
$|01\rangle$ ($|10\rangle$) computational state are plotted when
qubit 1 is biased at $f=0.508$. The coupling between $|1\rangle$
and $|2\rangle$ in the presence of an AC magnetic field is also
plotted. Qubit coupling equal to $\tau$ (same for both qubits) was
assumed in the calculation.}\label{Elements}
\end{figure}

Besides finding the appropriate bias points such that the middle
two eigenstates of the coupled system are very similar to the
middle two computational states, one must also verify that the
coupling between these states in the presence of an oscillating
magnetic field is non-zero.  The results of these calculations are
shown in figure \ref{Elements}.  The bias point of qubit 1, $f_1$,
must be chosen to be far from 1/2, but not too far. In these
calculations we take $f_1=0.508$. In the figure we see that when
qubit 2 is biased at around $f_2=0.51$, the first two system
excited states are very similar to the middle two computational
states, with overlap elements of about 0.97. At this same bias
point one sees a Rabi matrix element of about 0.02, which is more
than sufficient for our purposes.

\begin{figure}
\includegraphics[height=3.0in]{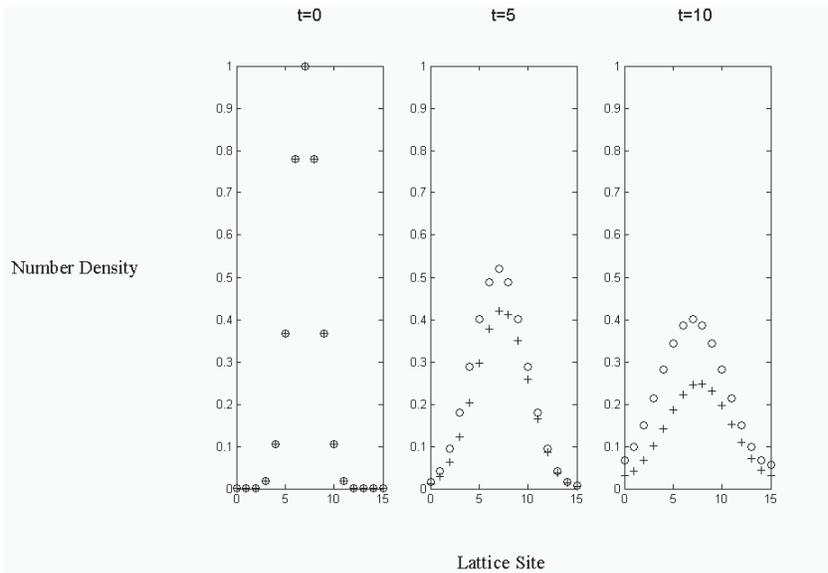}
\caption{The results of the FQLGA are simulated having accounted
for the approximate nature of the collision proposed in section
\ref{imp1}(+). Qubit coupling equal to $\tau$ (same for both
qubits) was assumed in the approximate swap simulation. The ideal
results of the FQLGA are also shown(o).} \label{ASB}
\end{figure}

This approximate swap has been incorporated into simulation of the
FQLGA for the 1D diffusion equation and the results are pictured
in figure \ref{ASB}. Snapshots of three different times have been
shown, for both an ideal simulation and one including the error
introduced due to the approximate collision. At time $t=0$ one can
see that we have initialized our fluid to a gaussian profile.
Later time steps of the ideal implementation show the expected
spreading due to diffusion, while conserving the total number of
particles. Increase in the diffusion constant of the approximate
collision when compared to the ideal simulation results from the
enhanced population in the $|00\rangle$, $|01\rangle$, and
$|10\rangle$ states relative to the $|11\rangle$ state due to
extra matrix elements in the approximate swap that couple the four
states.

Even with an ideal swap operator, an interesting timing issue
arises upon non-adiabatically switching the bias fields from the
initialization settings to the proper settings to do a Rabi
oscillation between the two middle product states. We first
illustrate this timing issue and then show how it can be made
negligible by making a larger ensemble measurement.

\begin{figure}
\includegraphics[height=3.0in]{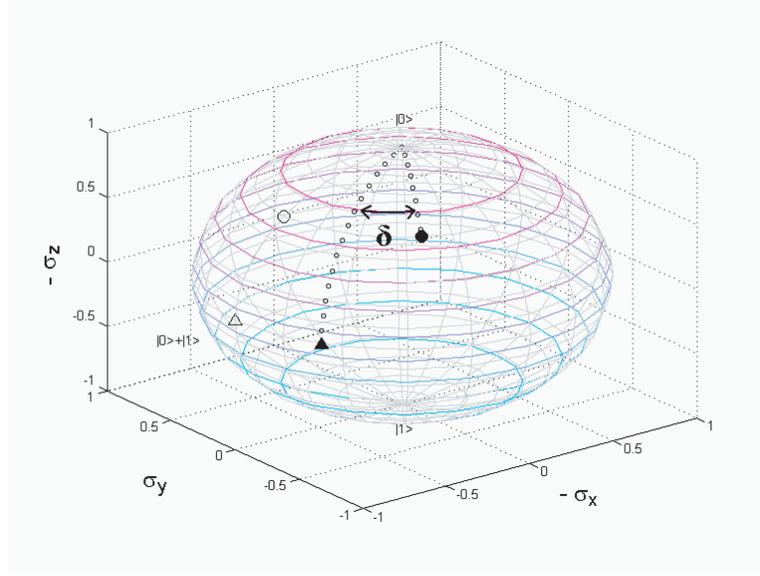}
\caption{The unfilled circle and triangle represent two typical
initialized states at one lattice point, both on the same north
pole to south pole geodesic, before their flux bias is changed to
perform the collision. The filled circle and triangle represent
the same states after imprecise bias changing has occurred.
Imprecisely timed Larmor precession introduces a random phase
difference $\delta$ between the two states.}\label{PhaseDiff}
\end{figure}

Once the applied fluxes are changed to those appropriate to
perform the approximate swap, the initialized states will most
likely not be eigenstates anymore, and hence will begin to precess
due to a time-independent perturbation. Assuming things can not be
accurately controlled at these timescales, one will have now
introduced a random phase difference between the two qubits due to
this Larmor precession. This effect is pictured in figure
\ref{PhaseDiff}. The states before the bias fields are switched
lie along the same geodesic. Upon changing the magnetic flux seen
by each qubit, the qubits begin to precess, out of phase.

\begin{figure}
\includegraphics[height=3.0in]{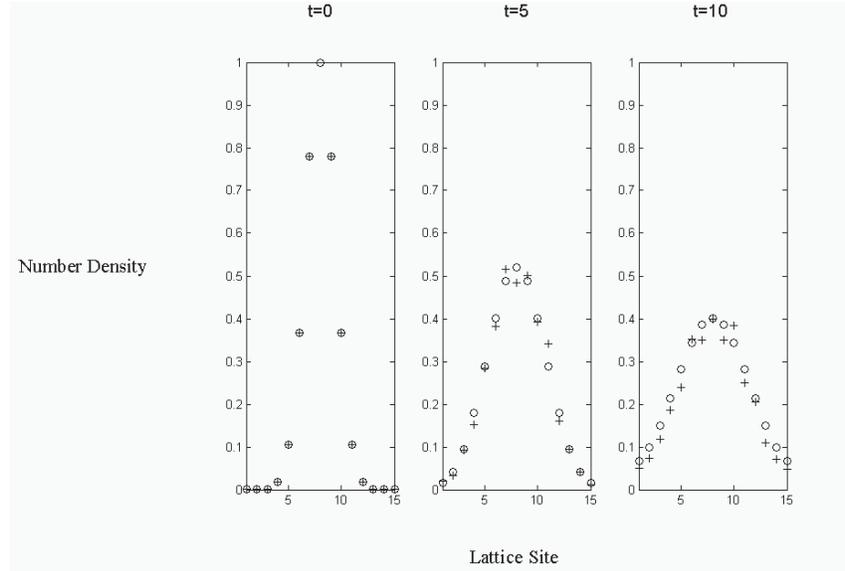}
\caption{The results of the FQLGA are simulated having accounted
for a random phase difference introduced before the collision for
each member of the measurement ensemble(+). The ideal results of
the FQLGA are also shown(o).}\label{RandomPhaseDiff}
\end{figure}

The effect of this phase difference $\delta$ on the algorithm will
be to alter the fraction of particles at each lattice site,
post-collision, that are ``moving" to the right and to the left.
The results of measuring the post-collision occupation
probabilities having accounted for a constant phase difference is
summarized by
     \begin{subequations}
     \begin{align}
P_1=P_{1,\ \delta=0} +
\gamma\sin(\delta)\label{first}\\
P_2=P_{2,\ \delta=0} - \gamma\sin(\delta).
\end{align}
\end{subequations}
The effect of this error on the simulation is effectively averaged
away when an ensemble is measured, since $\delta$ is randomly
different for each member of the ensemble.  These results are
shown in figure \ref{RandomPhaseDiff}. One can see small random
deviations from the ideal simulation that can be made
infinitesimally small by measuring a larger ensemble (an ensemble
average of 1000 repeated measurements was simulated here).

In summary, an initialization scheme has been developed that is
not available to qubits with only one term in their Hamiltonian.
This initialization scheme is limited only by the precision of the
current source used to create the magnetic field that biases the
qubit. The scalability of this scheme relative to those used in
NMRQC is an interesting question, but is not resolved here. The
collision implementation is also unique to qubits with multiple
term Hamiltonians, but the unitary transform implemented is unique
to the diffusion equation, and fortuitously simple. A collision
scheme that could be generalized to any unitary transformation
would be much more useful.

\subsection{Generalized NMRQC-like Implementation}\label{imp2}

Generalization of the above implementation to any fluid dynamics,
i.e., any unitary transformation, can be done in an analogous way
to NMRQC schemes.  Generalization of the collision transformation
consists of using a universal set of quantum computation gates,
and decomposing all transformations into a sequence of
these\cite{Chuang}. In NMRQC collision is performed by a sequence
of single qubit unitary transformations and coupled free
evolution. In this section we will begin by discussing the single
qubit rotations needed for a general decomposition, and briefly
mention the role they could play in initialization.  We will then
explore the free evolution of a coupled PC Qubit system, and then
show how to combine the single and coupled pulses to implement the
collision of the 1D FQLGA for the diffusion equation.

Single qubit transformations can most easily be achieved in a
rotating frame, since here the frequency of precession can be much
lower than the Larmor time scale.  For this implementation we will
only study the case where our qubit is biased at $f=1/2$.  This
discussion is easily generalized to any bias point, but the
mathematical notation can get quite cumbersome.  The Hamiltonian
of the PC Qubit in an applied AC field is
\begin{equation}
\hat{H}=w_o \hat{I}_x+g_o\cos(w_o t+\phi)\hat{I}_z
\end{equation}
where $w_o$ is the frequency of the applied field, $g_o$ is
proportional to the amplitude of the applied field, $\phi$ is the
phase of the applied field, and $\hat{I}_i=\hbar\hat{\sigma}_i/2$.

In the frame rotating about $\hat{I}_x$ this Hamiltonian becomes
\begin{equation}
\widetilde{\hat{H}}=\frac{1}{2}g_o[\cos(\phi)\hat{I}_z+\sin(\phi)\hat{I}_y].
\end{equation}
The quantum state will now precess in this frame about the axis
defined by $\phi$, with the angle through which the state has
precessed given by $\theta=g_o t/2$. It will be convenient to only
consider the set of two rotations defined by $\phi = 0$ and
$\pi/2$, which are rotations about $\hat{I}_z$ and $\hat{I}_y$
respectively.  These rotations, denoted as $R_z$($\theta$) and
$R_y$($\theta$) respectively, can be used in conjunction to bring
the qubit state to any point on the Bloch sphere in the rotating
frame.

\begin{figure}
\includegraphics[height=3.0in]{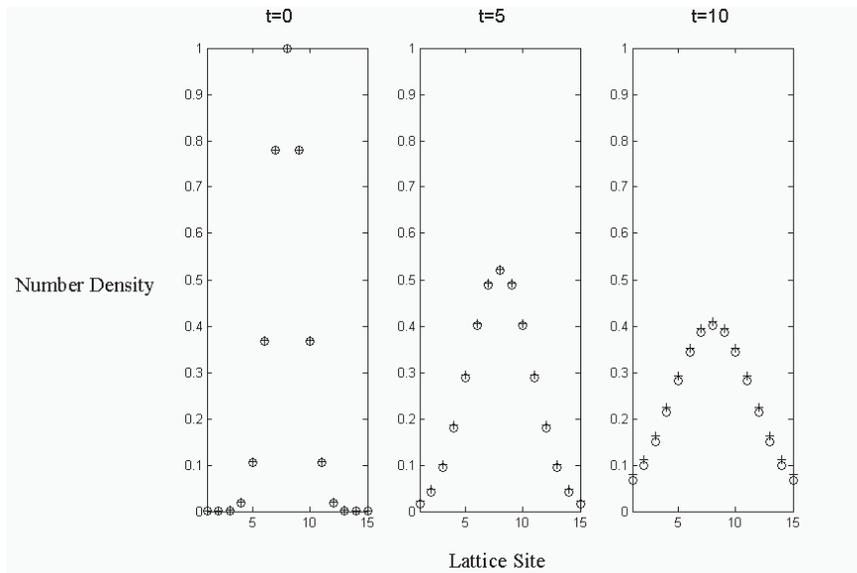}
\caption{The results of the FQLGA are simulated for NMR-like
single qubit pulse initialization, where errors arise from
initializing a coupled ground state that is not a product
state(+). The ideal results of the FQLGA are also shown(o).}
\label{CoupledInit}
\end{figure}

One can use these single qubit rotations not only as part of the
collision, but also for initializing, since they can bring the
qubit state to anywhere on the Bloch sphere. As already discussed
in section \ref{imp1}, the ground state of a coupled PC Qubit
system is not in general the product of single qubit ground
states. Thus, when initializing a qubit via $R_z$($\theta$) and
$R_y$($\theta$), one is not starting rotation from the single
qubit ground state. However, since the ground state is very close
to a product of single qubit ground states, this difference is
nearly negligible. In figure \ref{CoupledInit} we show the effects
of incorporating this error into the algorithm when the coupling
constant is taken to be 1/10 of the qubit resonant frequencies, a
rather exaggerated estimate since the coupling is usually much
smaller. The diffusion constant is decreased by this
approximation, due to the enhanced population in the $|11\rangle$
state relative to the $|00\rangle$ state from the coupling.


The other gate operation needed to form a universal set for
general decomposition is coupled free evolution. The first thing
to do is go to a co-rotating frame where one can have a coupled
Hamiltonian only, i.e., no single qubit terms, that is
time-independent.  In NMRQC this is done by going to the frame
where both qubits are rotated around the $\hat{z}$ axis. However,
since our coupling does not commute with our single qubit terms, a
different method will be used. For notational convenience only, we
consider the case where both qubits are biased at $f=1/2$, where
our Hamiltonian is
\begin{equation}
\hat{H}=w^{1}_o \hat{I}^{1}_x + w^{2}_o \hat{I}^{2}_x +
\frac{2\pi}{\hbar} J_{12}[\hat{I}^{1}_z\hat{I}^{2}_z].
\end{equation}

In the co-rotating frame where both qubits are rotating around the
$\hat{x}$ axis, one has the Hamiltonian
\begin{equation}
\hat{H}=\frac{\pi}{\hbar}
J_{12}[\hat{I}^{1}_z\hat{I}^{2}_z+\hat{I}^{1}_y\hat{I}^{2}_y]
\end{equation}
as long as $w^{1}_o = w^{2}_o$. This constraint of $w^{1}_o =
w^{2}_o$ imposes limitations on some NMRQC initialization schemes
which use frequency selective initialization.

One can now rewrite the unitary collision transformation in the
following suggestive way:
\begin{equation}
\sqrt{swap}=\exp[-i\frac{\pi}{8}(\hat{\sigma}^{1}_z\hat{\sigma}^{2}_z+\hat{\sigma}^{1}_y\hat{\sigma}^{2}_y)]
\exp[-i\frac{\pi}{8}\hat{\sigma}^{1}_x\hat{\sigma}^{2}_x]
\end{equation}
The first term is just free evolution in the co-rotating frame.
The second term can be written as:
\begin{equation}
\exp[-i\frac{\pi}{8}\hat{\sigma}^{1}_x\hat{\sigma}^{2}_x]=
R^{1}_y(\frac{\pi}{2})R^{2}_y(\frac{\pi}{2})
\exp[-i\frac{\pi}{8}\hat{\sigma}^{1}_z\hat{\sigma}^{2}_z] R^{1}_y
(-\frac{\pi}{2})R^{2}_y(-\frac{\pi}{2})
\end{equation}
where the middle term can be written as:
\begin{equation}
\exp[-i\frac{\pi}{8}\hat{\sigma}^{1}_z\hat{\sigma}^{2}_z]=
\exp[-i\frac{\pi}{8}(\hat{\sigma}^{1}_z\hat{\sigma}^{2}_z+\hat{\sigma}^{1}_y\hat{\sigma}^{2}_y)]R^{1}_z
(\pi)\exp[-i\frac{\pi}{8}(\hat{\sigma}^{1}_z\hat{\sigma}^{2}_z+\hat{\sigma}^{1}_y\hat{\sigma}^{2}_y)]
R^{1}_z (\pi)
\end{equation}
Hence one can perform a decomposition of the collision
transformation into a sequence of single qubit rotations and
coupled free evolution.

In summary, we have shown that the PC Qubit can implement the
unitary transform that performs collisions in the 1D FQLGA for the
diffusion equation by a single and coupled qubit evolution
decomposition.  The single qubit rotations were shown to be
feasible for qubit initialization as well, with a slight
approximation due to the coupled ground state that is not a
product state.  The coupled free evolution was seen to require
identical qubit frequencies over a lattice site, making
initialization a bit more challenging.

\section{Conclusions}

In this paper we have shown that the implementation of the FQLGA
for the 1D diffusion equation is feasible with PC Qubits.  We
began by considering the simplest scheme possible using the PC
Qubit.  This consisted of first initializing the qubits while
keeping them in their ground state, and then performing the
collision by quickly changing their flux bias points and then
performing a single $\pi/2$ pulse.  This initialization technique
could prove useful, but the way we have implemented the collision
is not easily generalized to other collisions.  We needed to
develop a more general collision scheme, and then see how we could
initialize in conjunction with that new scheme.

A more general collision transformation was then discussed by
decomposing the unitary matrix into a sequence of single qubit
rotations and coupled free evolution.  We first developed single
qubit rotations for the PC Qubit that could be used as part of the
collision decomposition as well as for initializing the occupation
probabilities.  The initialization was considered only approximate
due to the permanent non-commuting coupling between qubits. For
the coupled free evolution we saw that transforming to a rotating
frame analogously to NMRQC set a strong but feasible constraint on
the frequencies of our qubits. Ultimately one would like to remove
the constraint of equal frequencies, so that frequency-selective
initialization can be done analogously to the NMRQC
implementation, alongside the very general collision scheme. One
would then also need to account for initialization pulses rotating
states from a non-product ground state.

\begin{acknowledgements}
The authors would like to thank Debra Chen and Jeff Yepez for
valuable discussions.  This work is supported by the AFOSR/NM
grant FA 9550-04-1-0221.
\end{acknowledgements}


\end{document}